\documentclass[aps,prd,onecolumn,groupedaddress,showpacs,nofootinbib,amssymb]{revtex4}
\usepackage{graphicx,bm}
\usepackage[english]{babel}
\usepackage{amsmath}
\usepackage{amssymb}
\usepackage{amsfonts}

\begin{document}

\def\cL{{\cal L}}
\def\be{\begin{equation}}
\def\ee{\end{equation}}
\def\bea{\begin{eqnarray}}
\def\eea{\end{eqnarray}}
\def\beq{\begin{eqnarray}}
\def\eeq{\end{eqnarray}}
\def\tr{{\rm tr}\, }
\def\nn{\nonumber \\}
\def\e{{\rm e}}

\title{Covariant renormalizable gravity and its FRW cosmology}

\author{Shin'ichi Nojiri$^1$ and Sergei D. Odintsov$^2$\footnote{
Also at Center of Theor. Physics, TSPU, Tomsk}}
\affiliation{
$^1$Department of Physics, Nagoya University, Nagoya 464-8602, Japan\\
$^2$Instituci\`{o} Catalana de Recerca i Estudis Avan\c{c}ats (ICREA)
and Institut de Ciencies de l'Espai (IEEC-CSIC),
Campus UAB, Facultat de Ciencies, Torre C5-Par-2a pl, E-08193 Bellaterra
(Barcelona), Spain
}

\begin{abstract}

We consider the diffeomorphism invariant gravity coupled with the
ideal fluid in the non-standard way. The Lorentz-invariance of the
graviton propagator in such a theory considered as perturbation over flat
background turns out to be broken due to the non-standard coupling with the ideal fluid. 
As a result the behavior of
the propagator in the ultraviolet/infrared region indicates that some
versions of such theory are (super-)renormalizable ones (with appearance
of only physical transverse modes). The FRW cosmology in same cases may be
different from the one in General Relativity with the possible
power law inflationary stage.

\end{abstract}


\maketitle

Gravity theory belongs to special class among the four fundamental interactions 
since the corresponding quantum field theory has not been constructed. 
The main problem is that if we consider the perturbation from the flat background, 
which has a Lorentz invariance, by using General Relativity, there appear 
the non-renormalizable divergences coming from the ultraviolet region in momentum space.
The superstring theories seem to be an almost unique candidate for the theory
of quantum gravity. 
It is interesting how the ultraviolet behavior could be effectively modified in the
superstrings theory. 
If the behavior of the graviton propagator in the ultraviolet region could be changed 
from $1/k^2$ to $1/k^4$, the theory could become ultraviolet renormalizable.
Here $k^2$ is the square of the four momenta. Such a change, however, necessarily breaks
the unitarity, since the action must include the higher derivatives with respect to the
time coordinate as in case of higher derivative gravity (see
book \cite{ilb} for a general review).

The idea proposed in ref.\cite{Horava:2009uw} (for the extension and
cosmological applications, see \cite{HoravaFollow}) is to modify the ultraviolet behavior 
of the graviton propagator in Lorentz non-invariant way as $1/\left|\bm{k}\right|^{2z}$, 
where $\bm{k}$ is the spacial momenta and $z$ could be 2, 3 or larger integers. 
We define $z$ by the scaling properties of space and time
coordinates $\left(\bm{x},t\right)$ as follows,
\be
\label{Hrv0}
\bm{x}\to b\bm{x}\ ,\quad t\to b^z t\ .
\ee
When $z=3$, the theory is UV renormalizable.
Then in order to realize the Lorentz non-invariance, one introduces the
terms breaking the Lorentz invariance explicitly (or more precisely, breaking full
diffeomorphism invariance) by treating the temporal coordinate and the spacial coordinates
in an anisometric way.
Such model has the diffeomorphisms with respect only to the time coordinate $t$ and
spacial coordinates $\bm{x}$:
\be
\label{Hrv01}
\delta x^i=\zeta^i(t,\bm{x}),\qquad \delta t=f(t)\ .
\ee
Here $\zeta^i(t,\bm{x})$ and $f(t)$ are arbitrary functions.
Recently, several works\cite{Charmousis:2009tc} criticized the model of
ref.\cite{Horava:2009uw} (see also \cite{HoravaFollow}). 
They claimed that unphysical modes could appear in the theory. 
Such an appearance could occur due to the lack of the full diffeomorphism invariance. 
The local symmetry usually puts constraints on the system and only the physical modes 
appear as propagating modes.
However, since above theory has not full diffeomorphism invariance, we might not be able to 
obtain full constraints to restrict the possible modes to the physical modes. Then in order
to obtain consistent gravity theory, one usually needs the theory with full
diffeomorphism invariance. 
Due to the lack of the full diffeomorphism invariance, we cannot, for example, choose 
the lapse function $N$ to be unity if $N$ depends on the spacial coordinates. 
In order to avoid the problem, Ho\v{r}ava \cite{Horava:2009uw} has assumed that 
the lapse function $N$ only depends on the time coordinate but not on the spacial 
coordinate. Due to the assumption, we do not obtain the local Hamilton constraint but 
only global one which requires that only the spacial integration of the Hamiltonian should 
vanish. It is not so clear if such a weak constraint can exclude all the unphysical modes 
or not. The problem does not seems to be settled.

In this note, we propose Ho\v{r}ava-like gravity model with full diffeomorphism invariance.
When we consider the perturbations from the flat background, which has Lorentz invariance,
the Lorentz invariance of the propagator is dynamically broken by the non-standard
coupling with a perfect fluid. The obtained propagator behaves as $1/{\bm{k}}^{2z}$ with
$z=2,3,\cdots$ in the ultraviolet region and the model could be perturbatively power counting
(super-)renormalizable if $z>2$ .
Not as in the 
model \cite{Horava:2009uw}, where unphysical modes
might 
appear \cite{Charmousis:2009tc}, only physical transverse modes appear in our model
due to the full diffeomorphism invariance.
The FRW dynamics of our model is also discussed. When $z\geq 3$, the FRW
dynamics does not change
from the Einstein gravity but when $z=2$, we obtain the modified FRW dynamics, 
where power law inflation as in quintessence or phantom like model, could occur.

Let us start from the Lorentz invariant action, or the action with full
diffeomorphism invariance, where a perfect fluid couples with gravity in non-standard way,
but due to the coupling, the effective Lorentz non-invariant action can be obtained.
The fluid could not correspond to the usual fluid like, radiation, baryons, and dark matter, etc. 
As we discuss later, the non-standard fluid may come from the higher excited modes in string theories. 

The starting action is
\be
\label{Hrv1}
S = \int d^4 x \sqrt{-g} \left\{ \frac{R}{2\kappa^2} - \alpha \left( T^{\mu\nu} R_{\mu\nu}
+ \beta T R \right)^2 \right\}\ .
\ee
Here $T_{\mu\nu}$ is the energy-momentum tensor of the fluid.
The action (\ref{Hrv1}) is fully diffeomorphism invariant and belongs to
general class of modified gravities (for review, see \cite{review}).
We consider the perturbation from the flat background $g_{\mu\nu} = \eta_{\mu\nu} + h_{\mu\nu}$.
Then the curvatures have the following form:
\be
\label{Hrv2}
R_{\mu\nu} = \frac{1}{2}\left[ \partial_\mu \partial^\rho h_{\nu\rho}
+ \partial_\nu \partial^\rho h_{\mu\rho} - \partial_\rho \partial^\rho h_{\mu\nu}
 - \partial_\mu \partial_\nu \left( \eta^{\rho\sigma} h_{\rho\sigma} \right)\right]\ ,\quad
R = \partial^\mu \partial^\nu h_{\mu\nu} - \partial_\rho \partial^\rho
\left( \eta^{\rho\sigma} h_{\rho\sigma} \right)\ .
\ee
The following gauge condition is chosen:
\be
\label{Hrv3}
h_{tt} = h_{ti} = h_{it} = 0\ .
\ee
Then the curvatures  (\ref{Hrv3}) have the following form:
\bea
\label{Hrv4}
&& R_{tt} = - \frac{1}{2}\partial_t^2 \left(\delta^{ij} h_{ij} \right) \ ,\quad
R_{ij} = \frac{1}{2}\left\{ \partial_i \partial^k h_{jk} + \partial_j \partial^k h_{ik}
+ \partial_t ^2 h_{ij} - \partial_k \partial^k h_{ij} \right\}\ ,\nn
&& R = \partial^i \partial^j h_{ij} + \partial_t^2 \left(\delta^{ij} h_{ij} \right)
 - \partial_k \partial^k \left(\delta^{ij} h_{ij} \right) \ .
\eea
For the perfect fluid, the energy-momentum tensor in the flat background has the following form:
\be
\label{Hrv5}
T_{tt} = \rho \ ,\quad T_{ij} = p \delta_{ij} = w \rho \delta_{ij}\ .
\ee
Here $w$ is the equation of state (EoS) parameter.
Then one finds
\bea
\label{Hrv6}
&& T^{\mu\nu} R_{\mu\nu} + \beta T R \nn
&& = \rho \left[ \left\{ - \frac{1}{2} + \frac{w}{2} + \left( - 1 + 3w \right) \beta \right\}
\partial_t^2 \left(\delta^{ij} h_{ij} \right)
+ \left( w - \beta + 3w \beta \right) \partial^i \partial^j h_{ij}
+ \left( - w + \beta - 3w \beta \right)
\partial_k \partial^k \left(\delta^{ij} h_{ij} \right) \right]
\eea
If we choose
\be
\label{Hrv7}
\beta = - \frac{w-1}{2\left(3w - 1\right)}\ ,
\ee
the second term in the action (\ref{Hrv1}) has the following form
\be
\label{Hrv8}
\alpha \left( T^{\mu\nu} R_{\mu\nu} + \beta T R \right)^2
= \alpha \rho^2 \left(\frac{w}{2} + \frac{1}{2} \right)^2 \left\{  \partial^i \partial^j h_{ij}
 - \partial_k \partial^k \left(\delta^{ij} h_{ij} \right) \right\}^2\ ,
\ee
which does not contain the derivative with respect to $t$ and breaks the Lorentz invariance.
We now assume $\rho$ is almost constant.
Then in the ultraviolet region, where $\bm{k}$ is large, the second term in the action (\ref{Hrv1})
gives the propagator behaving as $1/\left| \bm{k} \right|^4$, 
which renders the ultraviolet behavior (compared with Eq.(1.4) in \cite{Horava:2009uw}).

Instead of (\ref{Hrv3}), one may choose covariant gauge condition although
we may need the ghost fields to keep the unitarity:
\be
\label{Hrv9}
\partial^\mu h_{\mu\nu} = 0\ \mbox{or}\ \partial^t h_{t\nu}= - \partial^i h_{i \nu}\ .
\ee
Then instead of (\ref{Hrv6}), it follows
\bea
\label{Hrv10}
&& T^{\mu\nu} R_{\mu\nu} + \beta T R \nn
&& = \rho \left[
\left\{ - \frac{1}{2} + \frac{w}{2} + \left( - 1 + 3w \right) \beta \right\}
\left\{\partial_t^2 \left(\delta^{ij} h_{ij} \right) + \partial_k \partial^k h_{tt} \right\}
+ \left( 1 + \beta - 3w \beta \right) \partial_t^2 h_{tt} \right. \nn
&& \left. \qquad + \left( - \frac{w\alpha}{2} - \beta + 3w \beta \right)
\partial_k \partial^k \left(\delta^{ij} h_{ij} \right) \right]
\eea
We should note the gauge condition (\ref{Hrv9}) gives $\partial^\mu \partial^\nu h_{\mu\nu} = 0$,
that is,
\be
\label{Hrv11}
\partial_t^2 h_{tt} = - \partial^i \partial^j h_{ij} - 2 \partial^i \partial^t h_{t \partial_i}
= \partial^i \partial^j h_{ij}\ .
\ee
Then if imposing the condition identical with (\ref{Hrv7}), we obtain (\ref{Hrv8}), again.

Note that the form (\ref{Hrv7}) indicates that the longitudinal mode does
not propagate but only the transverse mode propagates.
Not as in Ho\v{r}ava's model \cite{Horava:2009uw}, where unphysical modes
might appear \cite{Charmousis:2009tc},
only physical transverse modes appear in our model due to the full diffeomorphism invariance
in the action (\ref{Hrv1}).

There are two special cases in the choice of $w$: when $w=-1$, which
corresponds to the cosmological constant, one gets 
$T^{\mu\nu} R_{\mu\nu} + \beta T R=0$ and
therefore we do not obtain $1/\bm{k}^4$ behavior. When $w=1/3$, which corresponds
to the radiation or conformal matter, $\beta$ diverges and therefore there
is no solution.

The apparent breakdown of the Lorentz symmetry in (\ref{Hrv8}) occurs
due to the coupling with the perfect fluid. 
The action (\ref{Hrv1}) is invariant under the diffeomorphism 
in four dimensions and the energy-momentum tensor $T_{\mu\nu}$ of the non-standard fluid 
in the action should transform as a tensor under the diffeomorphism. 
The existence of the fluid, however, effectively breaks the Lorentz symmetry, 
which is the equivalence between the different inertial frames of reference.
Note that the expression (\ref{Hrv5}) is correct in the reference frame where the fluid does not flow,
or the velocity of the fluid vanishes. In other reference frames, there
appear non-vanishing $T_{it}=T_{ti}$ components and there could appear the derivative 
with respect to time, in general. 
This situation is a little bit similar to the vacuum solution of the Einstein gravity, which is, 
of course, fully diffeomorphism invariant, but the obtained Schwarzschild solution is 
not Lorentz invariant. 
This is a kind of spontaneous breakdown of the symmetry, which is popular in the quantum field theory. 
The spontaneous breakdown of the symmetry is the phenomenon that the symmetry in the action is not realized 
in the vacuum. In the field theory, 
we consider the quantum theory starting from the vacuum with the broken symmetry. 
Similarly we may consider the quantum theory of the gravity from the background where the non-standard 
fluid exists and breaks the Lorentz symmetry. 
Then we may obtain (power-counting) renormalizable quantum gravity. 

In the arguments after (\ref{Hrv2}), we have assumed the flat background
but the arguments could be generalized in the curved background: Even in the curved 
spacetime, due to the principle of the general relativity, we can always choose the local 
Lorentz frame. The local Lorentz frame has (local) Lorentz symmetry. Even in the local Lorentz 
frame, the perfect fluid might flow and $T_{it}=T_{ti}$ components might not vanish. 
By boosting the frame, which is the 
(local)
Lorentz transformation, we have a special local Lorentz frame, where the fluid does not flow.
In the Lorentz frame, we can use the above arguments and find there is no breakdown
of the unitarity. Conversely, in a general coordinate frame, $T^{\mu\nu} R_{\mu\nu} + \beta T R$
can have a derivative with respect to time.

We should also note that the action (\ref{Hrv1}) admits the flat spacetime solution although there is
a perfect fluid expressed by $T_{\mu\nu}$. Especially as a vacuum solution, there are Schwarzschild
and Kerr black hole solutions.

The action  (\ref{Hrv1}) gives $z=2$ theory.
In order that the theory could be ultra-violet power counting renormalizable in $3+1$
dimensions, we need $z=3$ theory. In order to obtain such a theory we note that,
for any scalar quantity $\Phi$, 
the explicit forms of the covariant derivatives have the following form:
\be
\label{Hrv12}
T^{\mu\nu}\nabla_\mu \nabla_\nu \Phi + \gamma T \nabla^\rho \nabla_\rho \Phi
= \rho \left[ \left\{ - 1 + \left(-1 + 3w\right) \gamma \right\} g^{tt} \partial_t^2 \Phi
+ \left\{ w + \left(-1 + 3w\right) \gamma \right\} \partial_k \partial^k \Phi \right]\ ,
\ee
with a constant $\gamma$. Then if we choose
\be
\label{Hrv13}
\gamma = \frac{1}{3 w - 1}\ ,
\ee
one obtains
\be
\label{Hrv14}
T^{\mu\nu}\nabla_\mu \nabla_\nu \Phi + \gamma T \nabla^\rho \nabla_\rho \Phi
= \rho \left( w + 1 \right) \partial_k \partial^k \Phi \ ,
\ee
which does not contain the derivative with respect to time coordinate $t$.
This is true even if the coordinate frame is not local Lorentz frame.
The derivative with respect to time coordinate $t$ is not contained in any coordinate
frame, where the perfect fluid does not flow.
Then if we consider
\be
\label{Hrv15}
S = \int d^4 x \sqrt{-g} \left\{ \frac{R}{2\kappa^2} - \alpha \left( T^{\mu\nu} R_{\mu\nu}
+ \beta T R \right)
\left(T^{\mu\nu}\nabla_\mu \nabla_\nu + \gamma T \nabla^\rho \nabla_\rho\right)
\left( T^{\mu\nu} R_{\mu\nu} + \beta T R \right)
\right\}\ ,
\ee
with
\be
\label{Hrv16}
\beta = - \frac{w-1}{2\left(3w - 1\right)}\ ,\quad
\gamma = \frac{1}{3 w - 1}\ ,
\ee
we obtain $z=3$ theory, which is renormalizable.
If one considers
\be
\label{Hrv17}
S = \int d^4 x \sqrt{-g} \left[ \frac{R}{2\kappa^2} - \alpha \left\{
\left(T^{\mu\nu}\nabla_\mu \nabla_\nu + \gamma T \nabla^\rho \nabla_\rho\right)
\left( T^{\mu\nu} R_{\mu\nu} + \beta T R \right) \right\}^2 \right]\ ,
\ee
we obtain $z=4$ theory, which is super-renormalizable.
In general, for the case
\be
\label{Hrv18}
S = \int d^4 x \sqrt{-g} \left[ \frac{R}{2\kappa^2} - \alpha \left\{
\left(T^{\mu\nu}\nabla_\mu \nabla_\nu + \gamma T \nabla^\rho \nabla_\rho\right)^n
\left( T^{\mu\nu} R_{\mu\nu} + \beta T R \right) \right\}^2 \right]\ ,
\ee
with a constant $n$, we obtain $z = 2 n + 2$ theory.
Usually $n$ should be an integer but in general, we may consider
pseudo-local differential
operator $\left(T^{\mu\nu}\nabla_\mu \nabla_\nu + \gamma T \nabla^\rho \nabla_\rho\right)^n$
with non-integer $n$ (e.g. $n=1/2,3/2$ etc.).
We should also note that there are special cases, that is,
$w=-1$ and $w=1/3$.

The second terms in the actions (\ref{Hrv1}), (\ref{Hrv15}), (\ref{Hrv17}), and
(\ref{Hrv18}), which effectively break the Lorentz symmetry, are relevant only in the high
energy/UV region since they contain higher derivative terms. In the IR
region, these terms do not dominate and the usual Einstein gravity follows as a limit.

Forgetting the renormalizability and regarding the action (\ref{Hrv18}) to
be an effective action, we may consider negative $n$ case. In such a case,
the action (\ref{Hrv18}) represents the non-local theory.

We now briefly consider the cosmology. Starting from the FRW background 
with flat spacial part,
\be
\label{Hrv19}
ds^2 = - dt^2 + a(t)^2 \sum_{i=1,2,3} \left(dx^i\right)^2\ ,
\ee
it is easy to see that $T^{\mu\nu}\nabla_\mu \nabla_\nu + \gamma T
\nabla^\rho \nabla_\rho$ only contains
the derivative with respect to spacial coordinates. The extra terms in
the actions (\ref{Hrv15}), (\ref{Hrv17}) and (\ref{Hrv18}),
which correspond to $z\geq 3$ case, vanishes 
if the metric of the spacial part is flat 
and we obtain the same FRW solutions as those
in the Einstein gravity. The extra terms, however, could modify the perturbations from the
FRW solution.
There is, however, an exception corresponding to the action (\ref{Hrv1}), that is, $z=2$ case.
The $z=2$ model does not give power-counting renormalizable theory but it might express some
effective theory. For example, we may consider the following type of
very general action:
\be
\label{Hrv18b}
S = \int d^4 x \sqrt{-g} \left[ \frac{R}{2\kappa^2} - \sum_{n=0}^N \alpha_n \left\{
\left(T^{\mu\nu}\nabla_\mu \nabla_\nu + \gamma T \nabla^\rho \nabla_\rho\right)^n
\left( T^{\mu\nu} R_{\mu\nu} + \beta T R \right) \right\}^2 \right]\ .
\ee
Since the term with $n=N$ dominates in the very high energy region, we
have $z=2N+2$ theory, which is power-counting renormalizable.
In  low energy region, only the Einstein-Hilbert term dominates.
However, in an intermediate energy region,
$n=0$ term could dominate and effectively  the action (\ref{Hrv1})
follows.
For the action (\ref{Hrv1}), the extra term gives the modification from the Einstein
gravity since the FRW universe is not the local Lorentz frame.
In order to obtain the equation corresponding to the FRW equation, we assume the following
form of the metric
\be
\label{Hrv19c}
ds^2 = - \e^{2b(t)}dt^2 + a(t)^2 \sum_{i=1,2,3} \left(dx^i\right)^2\ .
\ee
Then the action (\ref{Hrv1}) has the following form:
\be
\label{Hrv19d}
S = \int d^4 x a^3 \left[ \frac{\e^{-b}}{2\kappa^2} \left(6\dot H + 12 H^2 - 6\dot b H\right)
 - 9\alpha \rho^2 \left(1+w\right)^2 \e^{-3b} H^4 \right]\ .
\ee
The equation corresponding to the FRW equation can be obtained by putting $b=0$ after
the variation over $b$ and it has the following form:
\be
\label{Hrv19b}
\frac{3}{\kappa^2} H^2 = \frac{27 \alpha \rho^2 (1+w)^2}{2} H^4 + \rho_\mathrm{matter} \ .
\ee
Here $\rho_\mathrm{matter}$ is the usual matter energy density 
corresponding to radiation, baryons, and dark matter, etc. but not corresponding 
to the fluid appeared in the action (\ref{Hrv1}). 
In the situation that the contribution from the matter could be neglected,
there are two cases. One is trivial case $H=0$ and another is
\be
\label{Hrv20}
H^2 = \frac{2}{9\kappa^2 \alpha \left(1 + w \right)^2 \rho^2}\ .
\ee
This equation is consistent only if $\alpha$ is positive.
Since the fluid corresponding to $\rho$ has a constant EoS parameter, we find
$\rho = \rho_0 a^{-3(1+w)}$ with a constant $\rho_0$.
Then the solution of (\ref{Hrv20}) is given by
\be
\label{Hrv21}
a(t) \propto \left(t_0 - t\right)^{-1/3(1+w)}\ ,
\ee
which is phantom like solution even if $w>-1$.
Then there might occur the inflation due to this phantom like solution.
The second term in (\ref{Hrv19b}) is relevant only if the curvature is rather large
if we choose the parameter $\alpha$ properly.
Then if the universe started with not so large curvature, the term becomes irrelevant and
we obtain standard cosmology.

We now consider the case that $\alpha$ is negative.
In this case, the matter energy density $\rho_\mathrm{matter}$
in (\ref{Hrv19b}) can dominate if compared with the term coming from the
Einstein part as $\rho_\mathrm{matter} \gg \frac{3}{\kappa^2} H^2$ in the early universe.
In such a case, $\rho_\mathrm{matter}$ might express the energy density of the inflaton.
If the matter corresponding to $\rho_\mathrm{matter}$ has a constant EoS parameter
$w_\mathrm{matter}$, $\rho_\mathrm{matter}$ behaves as
$\rho_\mathrm{matter} = \rho_{\mathrm{matter}\,0} a^{-3(1+w_\mathrm{matter})}$
with a constant $\rho_{\mathrm{matter}\,0}$.
Then by neglecting the l.h.s. of (\ref{Hrv19b}), we find
\be
\label{Hrv22}
\frac{27 \alpha \rho_0^2  (1+w)^2 a^{-6(1+w)}}{2} H^4
+ \rho_{\mathrm{matter}\, 0} a^{-3(1+w_\mathrm{matter})} \sim 0\ ,
\ee
which shows
$H^2 \propto a^{-3(1+w_\mathrm{matter})/2 + 3(1+w)}=a^{3/2 - 3
w_\mathrm{matter}/2 + 3w}$.
Hence, the solution of (\ref{Hrv22}) is given by
\be
\label{Hrv23}
a(t) \propto t^{2/3 \left( 1 - w_\mathrm{matter} + 2w \right)}\ ,
\ee
if $1 - w_\mathrm{matter} + 2w > 0$.
On the other hand, if $1 - w_\mathrm{matter} + 2w < 0$, we obtain
\be
\label{Hrv24}
a(t) \propto \left(t_0 - t\right)^{2/3 \left( 1 - w_\mathrm{matter} + 2w \right)}\ .
\ee
Eq.(\ref{Hrv23}) describes quintessence-like inflation but
Eq.(\ref{Hrv24}) corresponds to the phantom-like one.
Then depending on the values of $w_\mathrm{matter}$ and $w$, both kinds of
inflation could be realized.

Instead of the perfect fluid in (\ref{Hrv5}), where the pressure $p$
is proportional to the energy density $\rho$, we can consider more general fluid, whose EoS is
given by $p=f(\rho)$. Here $f(\rho)$ is general non-linear function of
$\rho$.
The cancellation of the time derivatives (arguments around (\ref{Hrv6})-(\ref{Hrv8})), which
is necessary for the unitarity, however, only occurs when the EoS is linear equation.
Then for general perfect fluid with EoS $p=f(\rho)$, the cancellation does
not occur
or cannot be realized. The other types of the coupling should be
considered in such a case.

In summary,
we proposed gravity models with full diffeomorphism invariance, 
which has a renormalizable property as in Ho\v{r}ava's model of gravity. 
When we consider the perturbation from the flat background, which has Lorentz invariance,
the Lorentz invariance of the propagator is dynamically broken by the non-standard
coupling with the perfect fluid. As a result, the corresponding theory
may be  power counting (super-)renormalizable.
Moreover, only physical transverse modes appear in our model due to
the full diffeomorphism invariance.
The FRW dynamics of such model is studied. When $z\geq 3$, the FRW
cosmology does
not change from the Einstein gravity cosmology. When $z=2$, we obtain the
modified FRW
dynamics, where quintessence/phantom-like inflation may occur.

We now speculate what could be the fluid which appears in the action (\ref{Hrv1}). 
In order that the theory could be renormalizable, the fluid must be non-relativistic even in 
a very high energy regions, which is very unusual. In the string theory, however, there appear infinite 
number of heavy particles as excited states of string. As the energy becomes higher and higher, 
heavier and heavier particles are created by the quantum effects and the particles could make 
a non-relativistic fluid in any high energy region. Then this kind of fluid could naturally appear in the 
string theories. 

As a final remark, let us note that it would be of
interest to consider other dynamical scenarios of Lorentz invariance
breaking in such a way that this invariance is restored at current dark
energy
epoch.

\section*{Acknowledgments}

The work by S.N. is supported in part by Global
COE Program of Nagoya University provided by the Japan Society
for the Promotion of Science (G07).
The work by S.D.O. is supported in part by MICINN (Spain) projects
FIS2006-02842 and PIE2007-50I023 and by LRSS project N.2553.2008.2.

\end{document}